\documentclass[12pt]{iopart}
\usepackage{graphicx}
\usepackage{iopams}  

\begin{document}

\title{Complex-{$k$} modes of plasmonic chain waveguides}

\author{M. Yan}

\address{Department of Applied Physics,
 School of Engineering Sciences,
 KTH - Royal Institute of Technology \\
 Isafjordsgatan 22, Kista 16440, Sweden}
\ead{miya@kth.se}
\vspace{10pt}

\begin{abstract}
Nanoparticle chain waveguide based on negative-epsilon material is
investigated through a generic 3D finite-element Bloch-mode solver which derives
complex propagation constant ($k$). Our study starts
from waveguides made of non-dispersive material,
which not only singles out ``waveguide dispersion'' but also motivates search of new materials
to achieve guidance at unconventional wavelengths. Performances of
gold or silver chain waveguides are then
evaluated; a concise comparison of these two types of chain waveguides
has been previously missing. Beyond these singly-plasmonic chain
waveguides, we examine a hetero-plasmonic chain system
with interlacing gold and silver particles, inspired by a recent proposal; the
claimed enhanced energy transfer between gold particles appears to be
a one-sided view of its hybridized waveguiding behavior --- energy
transfer between silver particles worsens. Enabled by the versatile
numerical method, we also discuss effects of
inter-particle spacing, background medium, and
presence of a substrate. Our extensive analyses show that the
general route for reducing propagation loss of e.g. a gold chain
waveguide is to lower chain-mode frequency with a proper geometry
(e.g. smaller particle spacing)
and background material setting (e.g. high-permittivity background or
even foreign nanoparticles). In addition, the possibility of building mid-infrared chain waveguides
using doped silicon is commented based on numerical simulation.
\end{abstract}

%
%
%
%
%

\section{Introduction}

\begin{figure}[t]
\centering
\includegraphics[width=12cm]{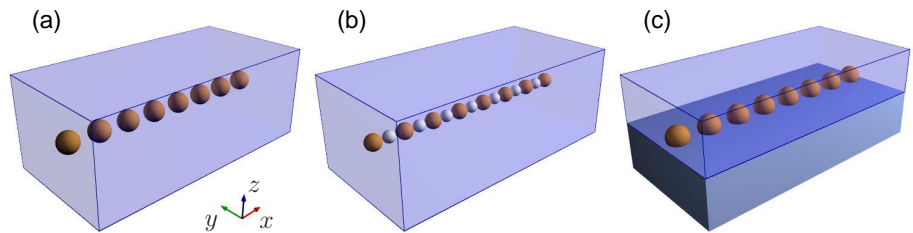}
\caption{Schematic diagrams of three types of chain waveguides. (a)
  Ideal chain waveguides based on negative-epsilon
  nanoparticles. Background medium (light blue region) can be a
  dielectric material or vacuum/air. (b) A chain waveguide consisting
  two types of particles (silver- and gold-colored). (c) A chain
  waveguide with dome-shaped particles sitting on some substrate.}
\label{fig:chains}
\end{figure}

Optical waveguide formed by a chain of metal nanoparticles (\emph{chain
  waveguide} in short) was proposed in \cite{Quinten:98} as an
alternative to axially invariant plasmonic waveguides for shrinking footprint of
photonic integrated circuits. Figure \ref{fig:chains}(a) illustrates a schematic picture of such
waveguide. Plasmonic chain waveguide relies on coupled
electromagnetic (EM) resonances in negative-epsilon particles to relay
optical wave. The particles are usually in deep-subwavelength scale
such that they can be treated as dipoles. That a chain of coupled
dipoles can channel EM power without any radiation
leakage contrary to stand-alone dipoles is in a sense
an extreme case where radiation of a dipole
is heavily affected by its environment. From waveguiding point of
view, it can be argued that such propagation of
EM wave in a periodic material system is fundamentally
indifferent from those all-dielectric periodic waveguides presented in
\cite{Fan1995,Yariv:99,Cheben:06,Cheben2018:SWG}, some of which were referred to as
coupled-resonator optical waveguides (CROWs) or sub-wavelength grating
waveguides. What make the waveguides under current investigation
distinct are the ``plasmonic'' nature of the modes as well as the
strong material dispersion, and thereby loss, usually associated with
negative-epsilon materials.

Plasmonic chain waveguide in its simple
setting has been studied mostly through theoretical
modeling~\cite{Quinten:98,Brongersma:2000:chain,Maier:2003:chainFDTD,Weber2004,Fung2007,Conforti2010,Guasoni2011,Dong2013,Compaijen2015}
and, in fewer occasions, through experiments~\cite{Krenn1999,Maier:2003:chainExperiment,Barrow:2014:EELS,Hanske:2014:templateAssembly,Crozier:07,Solis:2012:chainAssembled,Gur2018:DNAChain,Mayer:2019:templateAssemblyEELS}. There
are several ways to prepare chain waveguides. The standard approach is to
use electron-beam lithography (EBL) combined with lift-off process to pattern
metal nanoparticles~\cite{Krenn1999,Maier:2003:chainExperiment,Crozier:07}; the
method produces nanoparticles with residual surface
roughness and has difficulty in controlling feature
size below 10~nm. Alternatively, chemists can
routinely synthesize noble metal nanoparticles of preferred geometry 
and then let them self-organize on a pre-patterned (chemically or topographically) substrate
\cite{Solis:2012:chainAssembled,Hanske:2014:templateAssembly,Mayer:2019:templateAssemblyEELS,2019:Mayer:colloidalAssembly}. Especially,
self-assembly of colloidal particles on wrinkled elastomeric template
~\cite{Mayer:2019:templateAssemblyEELS,2019:Mayer:colloidalAssembly}
has produced chains with 20 and even more gold nanoparticles with
fairly homogeneous particle spacings ($<2$~nm). Very recently, based
on the DNA origami method\cite{Mirkin:DNAAssembly}, there
have been several work demonstrating well-structured plasmonic chains
with gap size precisely controlled at sub-10~nm
level\cite{Gur:2016:AuNPChain,Roller2017,Gur2018:DNAChain,Ye:2019:AuChain}. Nanotechnology
is seemingly on its way to make perfect chain structures. However,
a thorough knowledge of their ultimate waveguiding performance,
despite previous theoretical work, can hardly be readily looked
up.

Existing understanding of dipole-based
fundamental chain modes based on previous
work~\cite{Quinten:98,Brongersma:2000:chain,Maier:2003:chainFDTD,Weber2004,Fung2007,Conforti2010,Guasoni2011,Dong2013,Compaijen2015}
is summarized as follows. Unlike uniform
waveguides like optical fibers, the resonant nature of chain
waveguides determines that light propagation in such waveguides is
relatively narrow-band, with transmission frequencies decided roughly
by the resonant frequencies of individual nanoparticles. Depending on major polarization, there exist two types
of low-order chain modes: transverse mode (T mode) with electric field majorly
directed normal to waveguide axis ($x$ in this work, Fig.~\ref{fig:chains}) and longitudinal mode (L mode) with
electric field majorly directed along waveguide axis. T mode appears
in a degenerate pair since the waveguide is rotationally symmetric
with respect to the waveguide axis. The L mode is unique for chain
waveguides based on negative-epsilon materials; its
polarization is seemingly incompatible with the
transverse-EM (TEM) nature of light. Both T and L modes can be deep-subwavelength (yet it was not pointed out when
such tight guidance can be obtained). Due to almost inevitable
lossy nature of all negative-epsilon materials, guided modes in
chain waveguides tend to have a large propagation loss. Precise
calculation or demonstration of propagation loss, especially its wavelength dependence,
was very scarce. In \cite{Quinten:98}, it was found after geometrical optimization that
propagation length for a silver chain waveguide with a period of 75~nm in
vacuum is as short as $\sim 0.9\ \mu$m at near-ultraviolet (UV)
wavelength. In the latest experimental endeavor~\cite{Gur2018:DNAChain}, a gold chain
waveguide was shown to have a propagation loss (L mode) of 0.8~dB per
50~nm (which is about one period size), apparently too much for
information-transfer applications. Another recent
work \cite{Roller2017} demonstrated that adding a silver nanoparticle in between two gold
nanoparticles can increase energy transfer between the two gold
particles. The latter naturally opens up a question whether a hetero-plasmonic chain
waveguide, or one chain with interlacing gold and silver particles (Fig. \ref{fig:chains}b), can
potentially have a propagation length fulfilling practical application
needs.

To rigorously analyze a plasmonic chain waveguide is certainly not a trivial numerical
task. Previously dominant analysis methods include quasi-static dipole approximation with damping
correction at optical frequency \cite{Weber2004}, and
eigen-decomposition method in
\cite{Fung2007,Compaijen2015,Dong2013}. The dipole approximation
method derives complex frequencies as eigenvalues. For a certain mode,
the imaginary part of frequency denotes its lifetime. Wave number or
propagation constant $k$ of a mode is read from computed mode
pattern. Such mode analysis corresponds to a situation where a chain is uniformly
excited by a plane-wave EM field, whereas in waveguiding
problem one is more interested in finding a complex wave number $k$ with a
given real frequency (as laser sources usually have). The imaginary part
of $k$ denotes spatial decay of a mode along the waveguide. The
eigen-decomposition methods solves for scattering spectra
against frequency and wave number, both in real values. The width of a
band in frequency is interpreted as mode quality. The iterative
calculations of scattering spectra can however be a lengthy process. To the best of
our knowledge, there has been very few reports on direct solution of
complex-$k$ modes in chain waveguides with real frequency as
input. The only work doing such direct calculation is found in
\cite{Conforti2010,Guasoni2011}, where a semi-analytical method was
formulated using Mie theory with lowest-order approximation. All the
approaches mentioned above by default only handle
waveguide made of identical nanoparticles, usually
spherical. Furthermore, the methods usually have difficulty in including a
substrate. The current work aims to use an empirical mode solver based
on finite-element method (FEM) to study guidance properties of chain
waveguides in a broader context. The numerical mode solver is built
upon \cite{Davanco2007a} with incorporation of perfectly matched
layers (PMLs)\cite{Parisi:12} for solving complex wave number in a
restricted direction from a real frequency input. As the method is fully
numerical, it can handle any particle geometry or material composition in a unit cell,
including multiple dissimilar particles.

It shall be pointed out that most experimental and some theoretical work focused on
chains made of a few particles
(e.g. \cite{Willingham:11,Barrow:2014:EELS,Gur2018:DNAChain,Mayer:2019:templateAssemblyEELS}). Such
short chains, owing to reflections at two ends, support Fabry-P{\'e}rot resonances; consequently, a single
propagating mode in an originally infinitely periodic waveguide becomes
several modes appearing at distinct resonant frequencies. In other
words, a single Bloch mode becomes quantized into several modes. This
phenomenon happens particularly for the L mode in short gold chains~\cite{Willingham:11,Barrow:2014:EELS,Gur2018:DNAChain,Mayer:2019:templateAssemblyEELS},
where quantized L modes are referred to as L$_1$, L$_2$, L$_3$,
etc, depending on how many nodes in their field
profiles. Furthermore, the quantized modes are labeled as
\emph{super-radiant}, \emph{sub-radiant} or \emph{dark} modes,
depending on their overall dipole moments. It was hinted that
sub-radiant modes facilitate longer plasmon propagation in chain
waveguides because such modes do not suffer from high radiation
losses~\cite{Willingham:11}. Our focus in this work
is on chain waveguides with infinite lengths. An infinitely 
extending chain sustains one L mode and one pair of degenerate T modes originated from
coupling of particle dipole resonances. If their dispersion curves
(relationship between frequency and wave vector) stay below light line
of background medium, the modes do not suffer
from radiation loss at all. By knowing the dispersion
curves, we can then
design termination of a waveguide properly so as
to enhance resonance of a particular quantized L mode, either for
enhanced or reduced radiation loss. As a matter of fact,
for signal-transfer applications, reflection shall be minimized
through proper design of in- and out-coupling elements to a chain
waveguide; hence towards that purpose, focus on a short chain
and discussion on quantized L modes might be misleading.

Through this work we aim to give an accurate description of how chain waveguides
perform through rigorous Bloch-mode analysis. Several geometries and material
settings that are possibly producible by current nanofabrication
technologies are studied. The simulation results are cross-linked with
published experimental findings if possible. 
The paper is organized as follows. In Section II we recapitulate
general design rule for geometry selection in the context of plasmonic
nanoparticles involved, as well as other
considerations including convention for material permittivity and
details for the numerical method, etc. In Section III, the
investigation starts with waveguides
made of material with a real, wavelength-independent
negative-epsilon value. Such analysis with non-dispersive
material was exercised quite often in dielectric photonic-crystal
studies for capturing key modal characteristics. In such a
simplified setting, one clearly identifies the ``waveguide
dispersion'' (effect of geometry on modal dispersion, as commonly
referred to in fiber optics community),
instead of seeing mixed effect contributed also by variation in
permittivity. In the meantime, modal properties at
different frequencies can be related to distinct materials possessing
such permittivity values, which motivates us to look for new
negative-epsilon materials beyond the traditional noble
metals. Section IV then looks into
waveguides made of realistic dispersive materials, including silver and
gold for visible-light guidance; a side-by-side comparison of these
two types of chain waveguides was not reported before. Based on the
understanding, we evaluate performance of hetero-plasmonic 
waveguide and comment on what is really gained (and lost) with this
design. Under the same section, we also look into 
effect of change in homogeneous dielectric background and even more complicated 
environment --- non-spherical particles sitting on a solid substrate
(Fig. \ref{fig:chains}c). In addition, hinted by the geometry
dispersion obtained in Section III, we investigate prospect of using
doped semiconductors, an emerging plasmonic material for applications at mid-infrared (MIR)
frequencies \cite{Soref08:MIRPlasmonics,Ginn2011:SiPlasmonics,Karmer2015:DopedSiPlasmonics,Zhang2017:Plasmonics},
guiding light beyond visible and near-infrared (NIR)
wavelengths. Conclusion follows in the end.

\section{Geometry, material, and numerical considerations}
As mentioned in \cite{Fan1995}, for a certain 1D periodic waveguide,
there is a general upper frequency limit for guided mode. The limit is
set by the crossing point between light line in background medium
(i.e. $\omega=kc/n_b$) and the first Brillouin zone boundary
($k_{bz}=\pi/a$). $k$ is wave number; $c$ is speed of light; and $n_b$
is refractive index of background. Correspondingly,
vacuum wavelength for guided light in a certain chain waveguide has to be
\begin{equation}
  \lambda > 2n_ba.
  \nonumber
\end{equation}

The above equation can be used for determining an appropriate period
$a$, as the guiding wavelength $\lambda$ for a chain waveguide is
associated with the particle-plasmon resonance $\omega_\mathrm{par}$. Take silver as an
example --- silver has plasma frequency
$w_p=1.39\times~10^{16}\mathrm{rad/s}$ according to Drude-model
fitting \cite{PhysRevB.6.4370}. We consider in this study a background medium of refractive
index $n_b=\sqrt{\epsilon_b}=1.5$ (which can be
associated with many glass materials). One has
$\omega_\mathrm{par}=w_p/\sqrt{1+2\epsilon_b}=5.9\times 10^{15}$~rad/s in frequency
or $318$~nm in vacuum wavelength. For achieving light guidance at such
wavelength, one shall use a period less than
106~nm. In the following study, the reference structure has period
$a=80$~nm. Particle size, denoted by diameter $d$, in our primary focus is chosen as 60~nm. Such a
structure can accommodate guided modes, if existing, above vacuum wavelength
of 240~nm. The above analysis assumes that guided modes will
appear around $\omega_\mathrm{par}$, which is a rough
estimation. As will be shown in Section IV, guided chain modes are
dispersive, and can exist for silver-chain case at up to
$550$~nm in vacuum wavelength.

Throughout this work, we deploy time-harmonic convention of
$\exp(i\omega t)$. A lossy material shall have a relative permittivity
value $\epsilon=\epsilon_r+\epsilon_ii$ where the imaginary part
$\epsilon_i$ must be negative.
A forward-propagating wave along $+x$ then has
spatial dependence $\exp(-ikx)$, where the complex
wave number is expressed as $k=k_r+k_ii$. $k_r$ must be positive; $k_i$ is
negative for decaying field, and positive for amplifying field.

FEM based on Galerkin weak-form formulation of
magnetic-field wave equation \cite{Davanco2007a} is
implemented in COMSOL Multiphysics environment. One periodic unit is
used in waveguide's longitudinal (axial) direction, with four
transverse sides enclosed by a PML layer to emulate open-boundary
condition. Transverse simulation domain
has dimension $6.5\times6.5\mu\rm{m}^2$, including an outer PML with thickness at
250~nm. Mesh size ranges from 5 to 50~nm. Reflection symmetry in the
transverse domain is utilized to reduce
problem size. Discretization deploys quadratic vector element, which
results in a number of unknowns usually close to a half million. By restricting wave vector
(thereby also periodic boundary condition) along waveguide axis, one can find complex-$k$ 
propagating modes of a chain waveguide, given a real input frequency. Consequently, mode searching
must be carried out in a complex wave-number plane. Only guided modes
under the light line of background medium
are searched. These modes are evanescent in transverse domain,
therefore not contributing to crosstalk. There can exist
resonantly guided modes above the light line, which are leaky. These
leaky modes are blended in the continuum states of background
medium; it is difficult to get them filtered out by the current
method. High-order modes based on
multipolar particle resonances can also be calculated; they
are however not of focus in the current study.

\section{Non-dispersive metal waveguide}
\subsection{Lossless chain}

\begin{figure}[b]
\centering
\includegraphics[width=8.8cm]{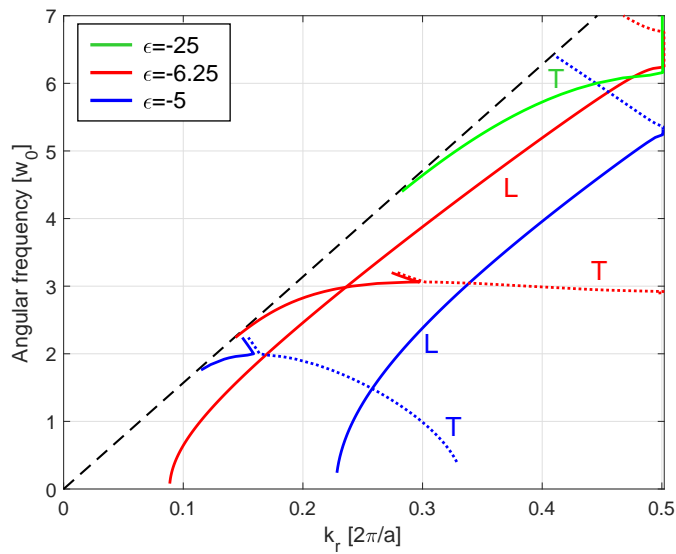}
\caption{Modal dispersion curves of chain waveguides with non-dispersive
  particle permittivity. Longitudinal (L) and transverse (T) modes are
  duly labeled for each dispersion curve. Geometry is fixed at $a=80$~nm and
  $d=60$~nm. $\epsilon_b=2.25$. Three
  particle permittivity values are studied, as shown in
  different colors. Dotted lines correspond to portions of dispersion
  curves with negative slopes. Note: dotted lines for
  $\epsilon_m=-5, -6.25$ are shifted by $+0.005\times 2\pi/a$ in
  $k_r$ to show the degenerate part. Dashed line is light line of the
  background medium.}
\label{fig:sm-d60a80-eps}
\end{figure}

Before we look into realistic plasmonic metals, we study mode
properties of chain waveguides made of fictitious non-dispersive
``metal'' particles. Three distinct permittivity values
($\epsilon_m=-5, -6.25, \mathrm{and} -25$) are considered, with
background medium $\epsilon_b=2.25$. The calculated bands for modes
identified (also called modal dispersion curves) are shown in
Fig.~\ref{fig:sm-d60a80-eps}. Note that when presenting mode dispersion curves of a periodic system made of
\emph{non-dispersive} material, one
often uses scaling law --- i.e. both frequency and
wave number are normalized against some length metric (most often
period of structure). In this work we consider
\emph{dispersive} metal materials; for ease of comparison, we choose to
normalize angular frequency against
$\omega_0=10^{15}$~rad/s. Wave number is normalized against $2\pi/a$
($a$ is chain period),
as conventionally done in photonic-crystal analyses. Wave number $k$ is in
general a complex number, which is even true for all-real
permittivity values used. Modes with frequency falling in
photonic band gaps carry such a complex $k$ value, with its imaginary
part denoting its degree of attenuation. As a common practice for
periodic systems, dispersion plots in this work only show
bands in the first Brillouin zone (BZ), with positive $k_r$.

The figure reveals that the chain waveguide
in general possesses both T and L modes over the examined frequency
range. The dispersions of modes can be quite 
peculiar as compared to conventional all-dielectric waveguides. The
dispersion curves of T modes are seemingly
stemming out of the light line of background medium at small $k_r$
values. This feature is shared by modes in dielectric waveguides,
and is fundamentally due to the fact that the T mode and the mode
corresponding to the light line have the same TEM
symmetry. Different from bands in dielectric waveguides, here in the
plasmonic case, the T-mode dispersion curves can reach a maximum in
frequency and thereafter exhibit a negative slope (represented by
dotted-line sections). Negative-slope dispersion curves were
previously considered for realizing light propagation with negative
group velocity (-GV), and the plateau point in a dispersion curve for achieving
``stopped'' light with zero GV (0GV). However,
as will be shown in Fig. \ref{fig:sm-d60a80_loss}, dispersion
close to 0GV point or with small GV in general is highly susceptible
to material loss. Note that at the 0GV
position, T modes from +GV and -GV sections converge and become a degenerate
pair; the pair branches out towards high
frequency with a steep slope, at the same
time their $k$ values turning into a complex-conjugate pair
(Fig. \ref{fig:sm-d60a80_loss}). This is a signature that the modes 
from both +GV and -GV sides enter into a photonic band gap regime,
as was also noticed in \cite{Conforti2010}. In contrast, for
non-dispersive dielectric photonic structures, extremities of a band
occurs at either center or edge of its BZ (or in general high-symmetry wavevector
points). The L mode in
Fig.~\ref{fig:sm-d60a80-eps} has a more intriguing dispersion
profile. Firstly, L mode
has no coupling with light line of the background medium. This is
fundamentally due to the unique polarization nature of the mode. Secondly,
if it does not reside in the radiation cone of the background medium
(which we did not look into), the L mode persists over a very broad
frequency range, even when frequency
approaches zero. This suggests such a nanostructred chain 
waveguide (especially with small particle permittivity value) can
potentially guide light with a very long wavelength. In addition, the
guided mode can have a very large effective mode index
$n_\mathrm{eff}=k_r/k_0$; therefore deep-subwavelength guidance can be ensured.

When the particle permittivity value increases negatively, in general, the
dispersion curves shift to higher frequencies. This can be explained
by the fact that the particles become more and more metallic at large
negative permittivity values; thereby EM field is
squeezed more out of the particles. Resonance is
sustained by effectively smaller and low-index space surrounding the particles,
hence the higher frequencies observed.

\begin{figure}[t]
\centering
\includegraphics[width=8cm]{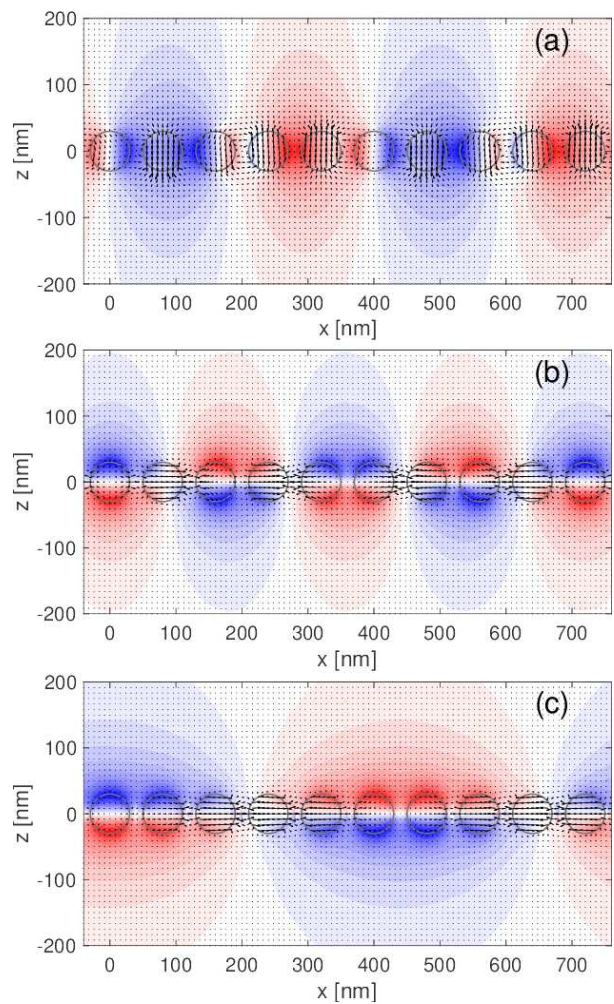}
\caption{Fields of lossless chain waveguides. $a=80$~nm,
  $d=60$~nm, $\epsilon_m=-6.25$, and $\epsilon_b=2.25$. (a,b) T and L modes at
  $\omega=2.8\omega_0$. (c) L mode at
  $\omega=0.3\omega_0$. The arrows denote displacement field,
  while the colormaps denote magnetic
  field component along $y$, i.e. H$_y$. Red and blue colors
  correspond to positive and negative field amplitudes, respectively.}
\label{fig:fieldsLossless}
\end{figure}

EM field patterns of three representative modes are presented in
Fig. \ref{fig:fieldsLossless}. Only cases for $\epsilon_m=-6.25$ are
plotted. For ease of interpretation, the fields
are constructed over multiple cells based on Floquet theorem to reveal
full-wavelength evolutions. Panels
(a) and (b) show the T and L modes at
$\omega=2.8w_0$, respectively. The frequency corresponds to $\lambda_0=673$~nm. The plots visualize clearly orientation
and coupling of particle dipoles, as well as wavelength of
the guided wave (thereof $n_\mathrm{eff}$ and mode confinement). For
the T mode in Fig. \ref{fig:fieldsLossless}(a), it has a major magnetic
field directed along $y$ direction, i.e. $H_y$ component (as shown by
the colormap in the figure). Whereas, the L mode in
Fig. \ref{fig:fieldsLossless}(b) has its magnetic field curling around
the waveguide axis. Comparatively, the L mode is somewhat similar to the
so-called transverse-magnetic mode in cylindrical optical
fibers. Figure \ref{fig:fieldsLossless}(c) shows an L mode at relatively a low frequency of $0.3\omega_0$,
corresponding to $\lambda_0=6.28~\mu$m. Deep-subwavelength guidance at
the MIR frequency is then achieved. We will discuss the possibility
of achieving such guidance using e.g. doped silicon in
Section IV.

L mode is heavily affected by particle spacing. A study
with a period of 100~nm (particle permittivity remains at -6.25) shows that
the L-mode band is markedly blue-shifted, by $\sim 21\%$ at $k_r=0.3[2\pi/a]$
(where $a=100$~nm), whereas the T-mode band is almost unchanged (increase in
frequency less than 2\% at the same wave number). The corresponding
dispersion curves can be found in ``Supplemental data''. Smaller particle
tends to increase band frequencies. This is confirmed with
a calculation of dispersion curves for a chain waveguide with
$a=80$~nm and $d=40$~nm, again with the same particle permittivity
(see also ``Supplemental data'').

\subsection{Lossy chain}

In the above sub-section, only lossless chain waveguides were
studied. As a matter of fact, material losses associated with
negative-epsilon materials are usually quite significant. Here we
look into the effect of adding an imaginary part to
the particle permittivity. Still, the considered particles will have
non-dispersive permittivity. Similar to the previous sub-section, the
focus is on a chain waveguide with $a=80$~nm, $d=60$~nm, and
$\epsilon_b=2.25$. A frequency-independent loss 
tangent of 0.01 is added to the material,
i.e. $\epsilon_m=\epsilon_r+\epsilon_ii=-6.25-0.0625i$. $\epsilon_i$
is negative for lossy material for the
time-harmonic convention used.

\begin{figure}[t]
\centering
\includegraphics[width=8cm]{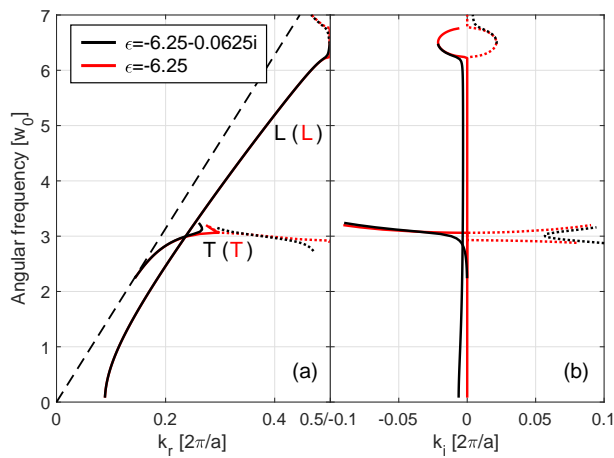}
\caption{Effect of material loss on modal dispersion curves of chain
  waveguides. $a=80$~nm, $d=60$~nm, and $\epsilon_b=2.25$. Metal
  particle has $\epsilon_r=-6.25$ with or without an imaginary part of
  $\epsilon_i=0.01\epsilon_r$. Bands corresponding to T and L modes
  are labeled.}
\label{fig:sm-d60a80_loss}
\end{figure}

In Fig. \ref{fig:sm-d60a80_loss}(a), dispersion curves obtained for
the lossless and lossy chain waveguides are compared. As mentioned
previously, even with all-real permittivity values, the mode solver
obtains modes with complex $k$,
more specifically for modes falling in photonic band gaps. When material loss
is introduced, modes at all frequencies carry complex $k$ values. The
imaginary part of $k$ ($k_i$) is 
plotted in Fig. \ref{fig:sm-d60a80_loss}(b). For a reason to be
clarified in the following paragraph, our trace of
dispersion usually stops when $k_i$ reaches $\sim$0.1. There are two
observations worth commenting for Fig.~\ref{fig:sm-d60a80_loss}. First (already pointed out in 
last sub-section), when loss is added, dispersion curve close to an originally
0GV point is highly affected. For the T mode, there are two 0GV
points, one at the plateau point of its dispersion curve and the other
at the BZ boundary. With material loss, the slopes of the dispersion curves at
those positions tend to increase rather than to approach zero. At the
same time, $k_i$ increases sharply. A direct consequence is that T
mode with -GV (dashed black line) becomes highly lossy. The more useful
T-mode dispersion curve (remaining +GV section, solid black
curve) bends back in $k_r$ before reaching the plateau
point, towards the originally degenerate mode-pair branch. The bending
starts sooner as material loss increases. This observation
suggests that small-GV modes are highly vulnerable to material losses; 0GV is simply not
possible. Similar finding on effect of material loss on GV was also
described in our previous investigation regarding a light absorbing
structure~\cite{Yan2013:absorber}. The second observation is that
dispersion-curve sections with negative slopes (dotted lines in Fig. \ref{fig:sm-d60a80_loss})
are associated with modes with amplifying amplitudes as they
propagation in $+x$ direction, which is manifested by their positive
$k_i$ values. In the current study, such amplifying mode propagation
applies to both the upper section of L band and the -GV section of the
T band. The effects and consequences of such amplifying
modes will be investigated in a separate study.

\begin{figure}[t]
\centering
\includegraphics[width=8.8cm]{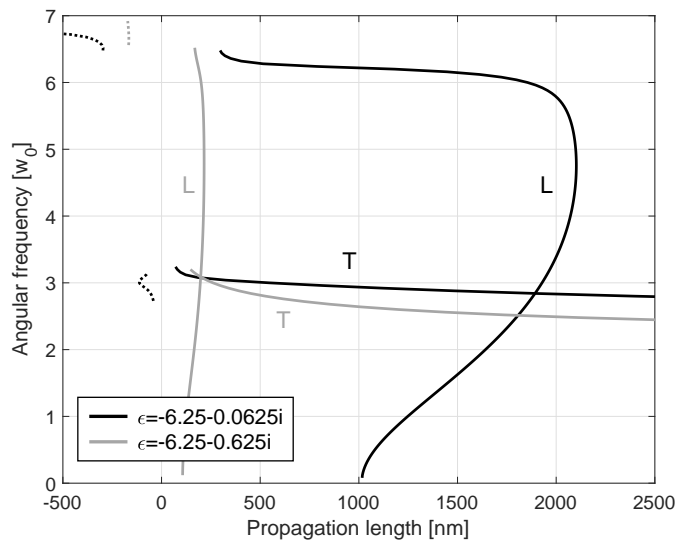}
\caption{Propagation length of chain-waveguide modes with particle material
  at two different loss levels. $a=80$~nm,
  $d=60$~nm, $\epsilon_b=2.25$. Metal particles have real permittivity
  $\epsilon_r=-6.25$, and a loss tangent of 0.1 or 0.01.}
\label{fig:sm-d60a80-pLength}
\end{figure}

Propagation length $L_p$ of a mode (distance for mode intensity decreasing
to its $1/e$) is calculated from imaginary part of wave number as
$L_p=-1/(2k_i)$, i.e. $L_p$ inversely proportional to $k_i$. When amplifying
modes are concerned, $k_i$ is positive and the distance in absolute value corresponds to
a mode's intensity increased to its $e$ times; we can refer to this length
particularly as amplification length. For reference,
a mode with $k_i=-0.1\left[\frac{2\pi}{a}\right]$ with $a=80$~nm has a
propagation length of 64~nm. This is already smaller than the
waveguide's period, rendering therefore such a waveguide almost useless at
the frequency considered. We plot in
Fig. \ref{fig:sm-d60a80-pLength} the propagation lengths for modes of
the waveguide at two loss tangents (0.1 and
0.01). The plot can be examined together with
Fig.~\ref{fig:sm-d60a80_loss}. Amplification lengths are plotted in
negative values. The T mode is found to be extremely
sensitive to frequency. Its $L_p$ can be relatively long
(beyond 1000~nm) in a short frequency range; however the long-$L_p$ modes are
quite close to the light line and therefore their modal confinements
are compromised. The L mode can sustain a consistent level of propagation
length over a large frequency range. Comparing the propagation lengths
at two loss levels, one finds the propagation length is roughly
inversely scaled to the imaginary part of metal
permittivity. Subwavelength mode confinement of L modes can be guaranteed as
its dispersion curve can stay quite far below the light line,
especially when small negative permittivity is used for particles (see
Fig.~\ref{fig:sm-d60a80-eps}). This observation 
hints that it would be interesting to look for new materials with small negative
permittivities for subwavelength light guiding at especially low frequencies.

\section{Real-metal waveguide}
Gold and silver are popular plasmonic materials. Gold has its
advantage of being chemically stable, while silver has lower loss at visible and
NIR frequencies. Experimental demonstrations of chain waveguides are
almost exclusively based on gold with the exception of
\cite{Roller2017} where a silver nanoparticle is incorporated between
two gold particles to enhance energy transfer. In this section we
evaluate chain waveguides made of these two classic metals. A rigorous comparison
of the two types of chain waveguides has not been presented
before. The data for silver and gold are taken from
\cite{PhysRevB.6.4370}; their analytical Drude models are not used since the
fittings are more relying on long-wavelength 
data and having large discrepancy at near-UV wavelength
regions. Experimentally measured permittivity data~\cite{PhysRevB.6.4370} show gold has a
much higher loss than silver at UV and visible
spectrum. For example, at $\lambda_0=550$~nm silver has $\epsilon_\mathrm{Ag}=-13-0.43i$ and gold has
$\epsilon_\mathrm{Ag}=-6-2.1i$, i.e 10 times difference in terms of
loss tangent. Later in this section, we will discuss
performance of chain waveguides at MIR frequencies, by using heavily-doped silicon. Such MIR
plasmonic material has its permittivity characterized by a Drude
model, as to be presented therein.

\subsection{Silver or gold chain}

\begin{figure}[b]
\centering
\includegraphics[width=8.8cm]{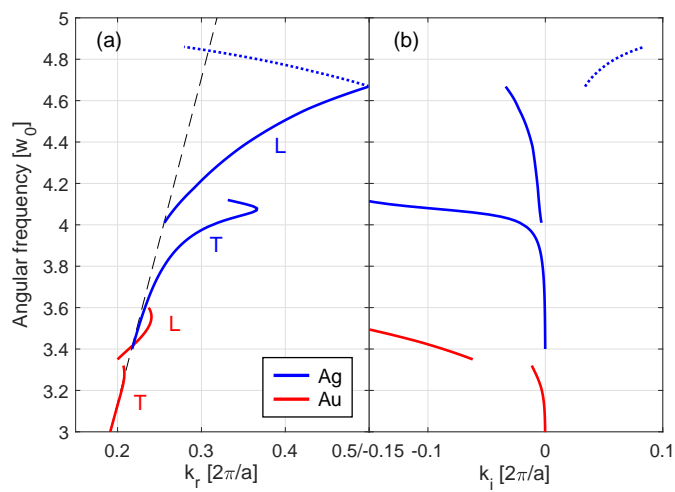}
\caption{Dispersion curves of modes guided in a silver (blue curves)
  or a gold (red curves) chain. $a=80$~nm, $d=60$~nm, and $\epsilon_b=2.25$. Both L
  and T modes are shown. (a) Real part of $k$; (b) imaginary part of $k$.}
\label{fig:disp_ag_au}
\end{figure}

In Fig.~\ref{fig:disp_ag_au} we lump together dispersion curves
calculated for a silver chain waveguide (blue curves) and those for a gold chain
waveguide (red curves), both with $a=80$~nm, $d=60$~nm, and
$\epsilon_b=2.25$. In general the dispersion curves for the silver chain waveguide appear at higher
frequencies (3.4--4.8~$\omega_0$, or 550--400~nm in
free-space wavelength, respectively), while the modes for gold
chain waveguides appear at 3--3.6~$\omega_0$ (630--520~nm
in free-space wavelength, respectively). This is in agreement with the
fact that the plasma frequency of silver is higher than that of
gold. As suggested by permittivity values, the silver chain waveguide suffers less propagation loss
compared to the gold counterpart. For the gold waveguide, the material
is so lossy that its T-mode dispersion curve tends to fold back as soon as it
stems out of the light line; as a result, the T-mode dispersion curve nearly laps over the
background light line and the mode is therefore quite delocalized. The L
mode has better confinement but it carries a $k_i$ around 0.1, hence
difficult to channel light across a distance larger than a period. The
silver waveguide is comparatively less lossy. At $\omega=3.91\omega_0$
and below, the T mode has $L_p>1000$~nm, with diverging $L_p$ as
frequency decreases (to $L_p=54 \mu$m at $\omega=3.4\omega_0$). The L
mode has $L_p>1000$~nm when $\omega<4.17\omega_0$ (until $~4\omega_0$ where
$L_p=1890$~nm). The results suggest that silver chain, especially its
T-mode, can be potentially used for channeling light at limited 
visible wavelength ranges. That said, it should be pointed out that
not all guidance is subwavelength; modes with long $L_p$ can be due to that
they stay close to the light line, therefore with their fields extended
in the background medium. One has to make a careful
comparison to other types of waveguides, plasmonic or not, in terms of some figure of
merit based on loss and mode confinement. The L mode tends to have
deep-subwavelength guidance. However, it has to be used with extreme care, since
it only achieves larger than 1000~nm propagation in a less than 20~nm wavelength window,
which can be further subject to experimental imperfections.

It was pointed out in \cite{Willingham:11} that smaller inter-particle
separation can increase plasmonic coupling and therefore increase
propagation length of chain mode. Extremely small separation of
$\sim$2~nm was used in a very recent
demonstration\cite{Gur2018:DNAChain}, and the work claimed a
L-mode propagation length of 190~nm for a gold chain. In comparison,
the gold chain presented in Fig. \ref{fig:disp_ag_au}, which differs
predominately with a larger particle separation of 10~nm, can hardly
propagate its L mode over one period. To verify the critical role
played by inter-particle distance, here a gold chain with $a=44$~nm,
$d=42$ (hence 2~nm gap), $\epsilon_b=2.25$ was modeled with our FEM
solver. This structure is very close to what was experimentally characterized in
 \cite{Gur2018:DNAChain}. In Section III, we concluded that a
 decrease in gap size will decrease mode frequencies (especially L
 mode), and a decrease in particle size will increase mode
 frequencies. Here the dominant effect is owing to the extremely small
 gap size. If we compare modal curves of the current structure to
 those of the gold chain in Fig.~\ref{fig:disp_ag_au}, its T-mode dispersion curve 
 remains almost unchanged, whereas its L-mode dispersion curve down-shifts
 to 2.65--3.15~$\omega_0$ (710--600~nm in wavelength). Propagation
 length is found to be as high as 800~nm at 710~nm wavelength, which
 is reduced to tens of nanometers at short-wavelength end. Still, this
 propagation length is in average seven times better than that of the
 gold chain in Fig.~\ref{fig:disp_ag_au}. Our numerical finding
in general supports the experimental observation of over 350~nm energy
transfer in the gold chain presented in \cite{Gur2018:DNAChain}, although the
dispersive nature of the modes was not mentioned in
\cite{Gur2018:DNAChain}. From our numerical investigation, the
deciding factor for achieving longer propagation length (compared to
the gold chain in Fig.~\ref{fig:disp_ag_au}) is that the L mode
appears at a lower frequency where gold material has less absorption.

\subsection{Hetero-plasmonic chain}
In Fig. \ref{fig:chains}(c), a heterogeneous chain waveguide is
illustrated, where two types of metal spheres are interleaved.
Such design was motivated by \cite{Roller2017}. It was proposed in
\cite{Roller2017} that by inserting a silver nanoparticle in between two
gold nanoparticles one can facilitate
more efficient energy transfer between two gold particles. The
argument was tested with plasmonic trimer structures, which were
realized by a delicate DNA-based self-assembly procedure (method also
used in \cite{Gur2018:DNAChain}). The loss reduction in energy
transfer was argued through reduced bandwidth of observed dark-field scattering
spectrum. Energy transfer across a longer chain beyond the trimer
structure was not discussed. Here we extend the idea to a
hetero-plasmonic chain waveguide, and numerically check its modal
properties including propagation length. It is worth noting that, in a \emph{finite} Au-Ag-Au trimer
structure, it was the energy transfer process between two \emph{gold
  particles} that was examined and was found to be more efficient
compared to two directly coupled gold particles; in an
\emph{infinite} hetero-plasmonic waveguide, energy transfer can happen
through hopping between \emph{silver particles}, which also has to be considered.

\begin{figure}[t]
\centering
\includegraphics[width=8.5cm]{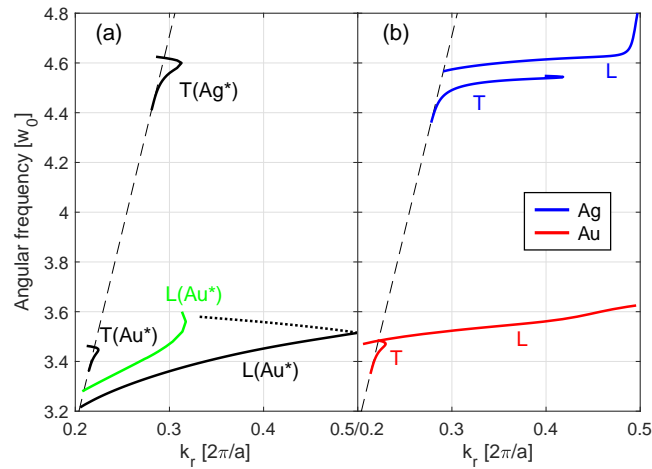}
\caption{(a) Modal dispersion curves of a hetero-plasmonic chain waveguide. $a=80$~nm,
  $d_{\textrm{Au}}=40$~nm, $d_{\textrm{Ag}}=30$~nm, and
  $\epsilon_b=2.25$. The black curves are calculated with imaginary part of gold and silver permittivity values
reduced by a factor of ten. The green curve is the only L mode
remaining when the metals' losses take their realistic values. (b) Modal
dispersion curves of two chain waveguides. One is gold chain waveguide
with $a=80$~nm, $d=40$~nm, and $\epsilon_b=2.25$ (red curves). The other is silver
chain waveguide with $a=80$~nm, $d=30$~nm, and
$\epsilon_b=2.25$ (blue curves). In both cases, metals have their imaginary
permittivities reduced by a factor of ten. Dashed line is light line of background medium.}
\label{fig:s2im}
\end{figure}

\begin{figure}[t]
\centering
\includegraphics[width=8cm]{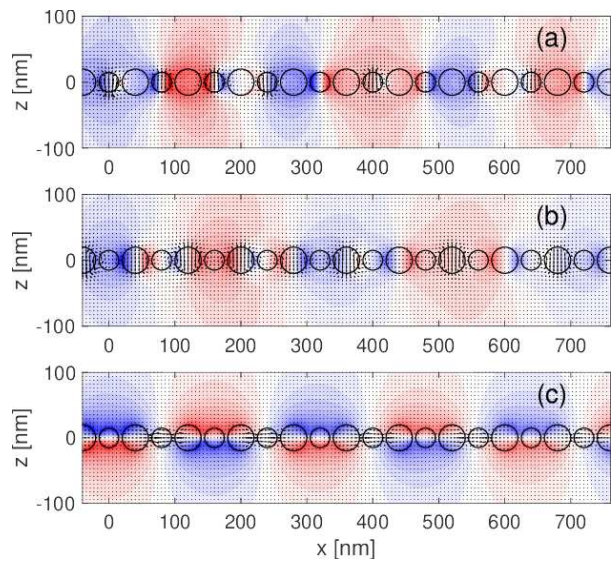}
\caption{Representative mode fields of the hetero-plasmonic chain waveguide. $a=80$~nm,
  $d_\mathrm{Au}=40$~nm, $d_\mathrm{Ag}=30$~nm,
  $\epsilon_b=-6.25$. (a) T mode at $\omega=4.55\omega_0$; (b) T mode
  at $\omega=3.44\omega_0$; (c) L mode at
  $\omega=3.30\omega_0$. The arrows denote displacement field,
  while the colormaps denote magnetic
  field component along $y$, i.e. H$_y$. Red and blue colors
  correspond to positive and negative field amplitudes,
  respectively. Imaginary parts of gold and silver permittivities are
  reduced by a factor of ten. Fields attenuate as a result of material
  absorption.}
\label{fig:fields_s2im}
\end{figure}

The hetero-plasmonic chain waveguide to be studied comprises of simultaneously
silver and gold nanoparticles. Our FEM analysis uses a supercell with a
period of 80~nm, including a gold particle of diameter 40~nm and a silver
particle of diameter 30~nm (thereof a gap size of
5~nm). $\epsilon_b=2.25$. The geometrical parameters as well as the
background permittivity
are very close to those studied in \cite{Roller2017}, where
gold-to-gold distance was 78~nm and background was assumed to have
$\epsilon_b=2.15$. It turns out that the hetero-plasmonic waveguide
under study is a very lossy waveguide --- key modal features are
killed by the presence of heavy material losses. In order to
better explain the guidance mechanism, we first calculated its dispersion
curves with imaginary parts of gold and silver permittivities
reduced by a factor of ten~\footnote{In fact, many modal dispersion
  calculations in the current work started with reduced or
  totally suppressed material losses. Such practice facilitates quick
  location of the bands.}. The resulted dispersion curves for the
guided modes are shown in 
Fig.~\ref{fig:s2im}(a) by the black curves. In order to know the nature
of the modes we plot three representative modal fields, one for each
dispersion curve, in
Fig.~\ref{fig:fields_s2im}. Examination of the mode
fields reveals that the two bands around $3.4\omega_0$ correspond to T
and L modes dominantly contributed by resonance in gold particles; and
the mode around $4.5\omega_0$ frequency is a T mode
dominantly contributed by resonance in silver particles.  The fact
that one type of particles is clearly in
resonance (dipolar) while the
other is not (acting somewhat like spacers) suggests that a hetero-plasmonic
waveguide can rather be treated as two superposed ``homo-plasmonic'' chain
waveguides. Each homo-plasmonic chain contain a single-type particles
with relatively large inter-particle spacings.

In Fig.~\ref{fig:s2im}(b) we present the dispersion curves of
gold/silver homo-plasmonic
chain waveguides, both with period $a=80$~nm but particle diameter
$d=40$~nm for the gold
case and $d=30$~nm for the silver case. The gold chain waveguide has T
and L bands around $3.5\omega_0$, corresponding well to the lower set of
modes in Fig.~\ref{fig:s2im}(a). The silver chain waveguide has T and L
bands around $4.5\omega_0$ with the T mode corresponding well to the upper
T band in Fig.~\ref{fig:s2im}(a). The L mode in silver chain waveguide
finds no counterpart in the hetero-plasmonic case. The lack of L mode
for the hetero-plasmonic chain waveguide is due to the fact that gold
is too lossy at the frequency. The effect of gold can be substantial
as the electric field of expected L mode has to majorly pass 
through the gold particles. If one examines modes' propagation lengths
(see ``Supplemental data''), one can conclude that: by
superimposing two homo-plasmonic chains into one hetero-plasmonic
chain waveguide, the chain modes supported by silver particles suffer
higher losses with the L mode disappearing completely; for the gold chain,
its T mode becomes a bit lossier, but the L mode can propagate much
longer. More specifically, the L mode of gold chain (Fig.~\ref{fig:s2im}b) has propagation
length ranging from tens of nanometers at its high-frequency end to 400~nm at
its low-frequency end ($3.470\omega_0$); the corresponding L mode in
hetero-plasmonic chain (Fig.~\ref{fig:s2im}a, solid black line) has 140~nm at its
high-frequency end and $\sim$1900~nm at the low-frequency end
($3.215\omega_0$). We refer to the gold-dominated L mode in the
hetero-plasmonic chain waveguide as L(Au$^*$) mode, and similarly for others.

In \cite{Roller2017}, the resonant modes were probed through
optical scattering spectra. If translated onto our
Fig.~\ref{fig:s2im}(a), their observed modes are within the light
cone, close to the $k_{r}=0$ axis if it was nearly
normal incidence). Their experiment recorded the dominant resonance
peak [corresponding to L(Au$^*$) mode] shifted ``from 549~nm
for the AuNP (gold nanoparticle) homodimer to 586~nm for the
heterotrimer structure''. These two peaks are respectively
$3.456\omega_0$ and $3.214\omega_0$ in frequency, which project quite
well to the curves that we have simulated. In \cite{Roller2017}, the
increase in mode quality was explained through concepts such as
``plasmonic hotspots'' and ``quasi-resonant virtual state'' of silver particle, however
we argue that the increase of mode quality is merely due to the fact
that inclusion of silver particles decreases the L(Au$^*$) mode to a
lower frequency where gold has smaller absorption loss. Losses of
silver material does not adversely affect mode quality of the L mode significantly,
since within that frequency gold has a loss tangent in
average ten times as large as that of silver.

When metals take their 100\% material losses, there is
only one gold-dominated L mode remaining, as shown by the green curve
in Fig.~\ref{fig:s2im}(a). Unfortunately its
propagation length is found to be less than 100nm.

The phenomenon of enhanced energy transfer is
geometry-dependent. Besides the structure examined above, we also looked
into a hetero-plasmonic waveguide with gold and silver
particles with identical diameter $d=60$~nm and a gap size of $20$~nm. The
gold-dominated L mode disappeared, possibly due to too large separation
between the gold particles.

Philosophically, it can be argued that that a hetero-plasmonic chain
waveguide is an improved gold-chain waveguide or a deteriorated
silver-chain waveguide.

We observed, through FEM analyses, that one
can insert dielectric (e.g. silicon or TiO$_2$) particles in a plasmonic
chain to create hybrid plasmonic-dielectric chain waveguide. Whether
such configuration has obvious advantage over singly-plasmonic chain
waveguide is subject to further investigation.

\subsection{Background medium and substrate}
The initial investigations on plasmonic chain waveguide used
air (or vacuum) as background medium~\cite{Quinten:98,Weber2004},
while we have so far focused on dielectric (glass) background. To see
the effect of background, we calculated dispersion curves of T and L
modes of a silver-in-air chain waveguide with $a=80$~nm and
$d=60$~nm (see ``Supplemental data''). Compared to the corresponding curves of the
silver-in-glass chain waveguide (Fig.~\ref{fig:disp_ag_au}), both
bands experience a blue-shift in frequency when air is used. The respective shifts are
roughly 26\% for the T mode and 18\% for the L mode, such that two bands are
crossing each other in the air-background case. Direct comparison of
propagation losses is not simple, as in two cases both modes are
dispersive: propagation loss is highly dependent on
frequency for each mode. In general, use of air background does not markedly
increase the propagation length of T mode; it actually shortens the
propagation length of L mode as a result of more flattened dispersion
curve of the mode. The L-mode band becomes more flat as a result of
effectively shortened gap size between the particles --- the chain
mode is therefore made of more localized gap plasmons with less
coupling among the ``hot spots''.

Actual nanofabrication of chain waveguides can lead to a background
that is more complex than a homogeneous medium. A more
realistic version is a chain made of nanoparticles sitting on top of a
dielectric substrate. Such geometry is common for chain waveguides
made from the template-assisted self-assembly or DNA-linking
process. Previously, the author and his colleagues have
reported experimental demonstration of arrayed gold nanoparticles
\cite{Wang2011d,doi:10.1021/nn2050032,Chen2014} on a dielectric
substrate based on light-induced rapid
thermal annealing of lithographically patterned rectangular metal
patches. The gold nanoparticles can have perfectly spherical shape, except with flattened bottom, similar to
droplets on a surface. The dome shapes are formed as a result of surface
tension when they are melted. A chain waveguide made of such metal
particles is schematically shown in
Fig.~\ref{fig:chains}(c). Inclusion of a substrate is problematic for
previously used theoretical
methods~\cite{Weber2004,Fung2007,Conforti2010,Guasoni2011,Dong2013,Compaijen2015,Barrow:2014:EELS},
but not for our FEM approach. We choose silver dome-shaped nanoparticles
for the following case study. In passing, we comment that one can have
other particle shapes like cubes and rods~\cite{2019:Mayer:colloidalAssembly}. The chain waveguide we are to study has the following parameters: $a=80$~nm,
$d=60$~nm, with a quarter of the sphere height truncated in the
bottom. The substrate has $\epsilon_s=2.25$. Rest of the background is air/vacuum.

\begin{figure}[t]
\centering
\includegraphics[width=8.8cm]{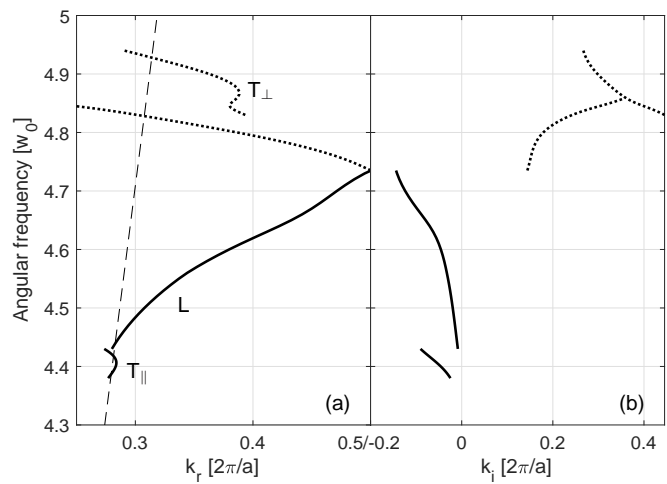}
\caption{Dispersion curves for modes guided by a dome-shaped
  silver-particle chain waveguide on a dielectric 
  substrate. $a=80$~nm, $d=60$~nm. The particles'
  spherical center is located from substrate by a distance of
  $d/4$. (a) Dispersion; (b) $k_i$. Dashed line is the
  light line of the substrate material.}
\label{fig:sos-d60a80}
\end{figure}

The dispersion curves of modes identified as well as their $k_i$
magnitudes are plotted in Fig.~\ref{fig:sos-d60a80}. It is known that when a substrate is
included, or when the particles depart from spherical symmetry, the
degeneracy of the T-mode pair  will be lifted. In other words, two
T-mode dispersion curves will appear, one with
major electric-field component directed perpendicular to the
substrate ($T_\perp$ mode) and the other with its major polarization parallel to the
substrate while perpendicular to waveguide axis ($T_\parallel$ mode). The two T-mode
dispersion curves are depicted in
Fig.~\ref{fig:sos-d60a80}(a). $T_\parallel$ mode appears at frequency
around $4.4\omega_0$, lower than that of the $T_\perp$ mode (around
frequency $4.9\omega_0$). The $T_\parallel$ mode has $L_p<250$~nm,
which is due to the mode is closely coupled to radiation mode in the
substrate material. The $T_\perp$ mode carries a
negative GV in general; its $k_i$ has a quite large magnitude in average. The
L mode exists in a relatively large frequency range with
$L_p<800$~nm. Overall, the propagation lengths obtained are not
encouraging for information transfer purposes. However, such
structure has an exposed surface which can be further functionalized for enhanced
nonlinear-optics and sensing applications, especially with the L mode exhibiting strong
gap-plasmon resonances.

We mention that the waveguide geometry has not been optimized. It
should be possible to tailor the particle shapes as well as to add
more thin substrate layers for further tuning of mode characteristics.

\subsection{MIR chain waveguide}
The dispersion diagram in Fig.~\ref{fig:sm-d60a80-eps} hints that if
one has a small negative permittivity ($-10<\epsilon_m<0$) at infrared wavelength
(e.g. $0.2\sim 0.5\omega_0$, or correspondingly $9.4\sim 3.8$~$\mu$m),
deep-subwavelength guidance of EM wave is possible. Such MIR waveguides
can be extremely interesting for integrated gas sensors. Here we
propose that one can possibly use heavily doped silicon to achieve the
desired negative permittivity and consquently a MIR chain
waveguide. It is already known that doped semiconductors can become
metallic at long EM wavelengths. Drude model can be used
for describing permittivity of doped silicon, as
$\epsilon(\omega)=\epsilon_\infty-\frac{\omega_p^2}{\omega^2+i\frac{\omega}{\tau}}$,
where $\epsilon_\infty$ is permittivity at high-frequency limit, 
$\omega_p$ is plasma frequency, and $\tau$ is collision time of free
carriers. Plasma frequency is calculated as
$\omega_p^2=\frac{Ne^2}{m^*\epsilon_0}$ with $N$ doping concentration,
$e$ elementary charge, $m^*$ effective mass of free carrier, and
$\epsilon_0$ vacuum permittivity. Following the numerical values
adopted in~\cite{Soref08:MIRPlasmonics}, we choose $n$-doped
silicon with $\epsilon_\infty=11.7$ and $m^*=0.272m_0$ ($m_0$ is electron
mass). Doping concentration chosen for the following
case study is at $N=2\times 10^{20}~\mathrm{cm}^{-3}$. Electron
collision time can calculated from measured carrier mobility as
$\mu=e\tau/m^*$. $\mu$ in~\cite{Soref08:MIRPlasmonics} was stated as
$\mu=50~\mathrm{cm}^2/\mathrm{(V\cdot s)}$ at the mentioned doping
concentration, which led to $\tau=7.73$~fs. We find that with this
collision time, Si permittivity has an imaginary part comparable to its
real part, which is too lossy for
making a waveguide. It was reported in~\cite{JACOBONI1977:Si} that
electrons' mobility can increase significantly at lower
temperature. At a doping concentration of $1.3\times
10^{17}~\mathrm{cm}^{-3}$, $\mu$ increases from $\sim
500~\mathrm{cm}^2/\mathrm{(V\cdot s)}$ at room temperature to over
$4000~\mathrm{cm}^2/\mathrm{(V\cdot s)}$ at 50~K. Although there were no
explicit experimental data for higher doping concentration scenario, here for a
theoretical exploration, we assume a certain low temperature can
increase $\mu$ by a factor of eight for the considered doping
concentration. Collision time increases correspondingly to
$\tau=387$~ps. The overall objective of the setting, as inspired by
Fig.~\ref{fig:sm-d60a80-eps}, is to obtain a
relatively small (less than 10) negative permittivity at MIR frequency
such that deep-subwavelength MIR guidance can be realized using a
chain waveguide made of such doped semiconductor materials. This idea
can extended naturally to chain waveguide design at even longer
wavelengths.

\begin{figure}[t]
\centering
\includegraphics[width=8.8cm]{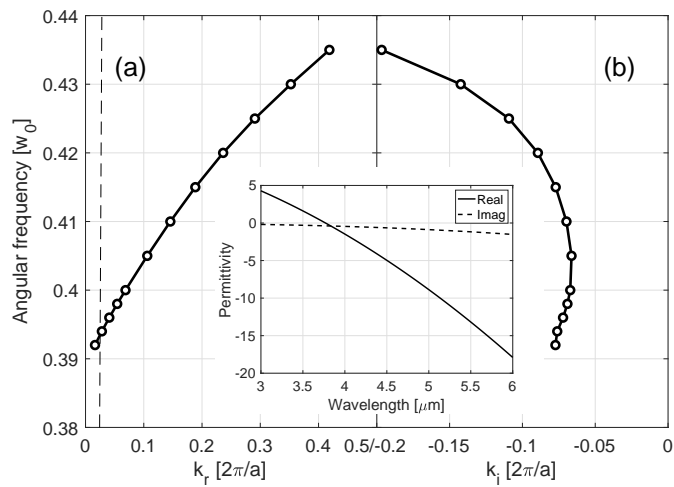}
\caption{Dispersion curve for L mode guided by MIR chain based on
  heavily doped silicon. $a=80$~nm, $d=60$~nm, and
  $\epsilon_b=2.25$. (a) Real part of $k$; (b) Imaginary part of
  $k$. Dashed line in (a) is background light line. Inset shows the permittivity of doped
  silicon around the guidance frequency range.}
\label{fig:sim-d60a80-si}
\end{figure}

By using the standard geometry in this work, i.e. $a=80$~nm
and $d=60$~nm, together with the above-mentioned particle material
setting, we find an L mode appearing around frequency $0.4\omega_0$,
as shown by Fig.~\ref{fig:sim-d60a80-si}a. Inset in
Fig.~\ref{fig:sim-d60a80-si} gives the real and imaginary parts of
silicon's permittivity at the relevant wavelength range. The background medium has
standard $\epsilon_b=2.25$. The imaginary part of the mode's
propagation constant (Fig.~\ref{fig:sim-d60a80-si}b) is however found to be
quite large in general. After conversion, the corresponding propagation length
hardly goes beyond 100~nm, with the longest at 96~nm when
$\omega=0.405\omega_0$.

\section{Discussion and Conclusion}
To conclude, a versatile FEM-based mode solver was formulated and used to
investigate modal properties of plasmonic chain waveguides in a few
varieties. The investigation started from chain waveguides made of
non-dispersive negative-epsilon materials. In this way,
``waveguide dispersion'' of both T and L modes were examined. The T
modes can have dispersion curves with a local maximum in the first
Brillouin zone; the curve section after the plateau point has negative
group velocity and amplifying amplitude when material loss is
present. While the T mode tends to exhibit flat band, the L mode tends
to span over a large frequency range. Moreover, the L 
mode can stay far below the light line of background medium, which hints
the possibility of achieving deep-subwavelength light guidance. Our analyses
then showed the performance of chain waveguides made of
realistic plasmonic metals. The silver chain waveguide, with the geometry and
background considered ($d=60$~nm, $a=80$~nm, $\epsilon_b$=2.25), can be promising for
achieving $>1$~$\mu$m light propagation, especially with its T
mode. However, as such a mode can be quite extended in background medium, one
has to motivate the use of a chain waveguide through some 
figure of merit based on loss and mode confinement. The L mode achieves $>1$~$\mu$m
propagation length (but not over 2~$\mu$m) in a very narrow wavelength
range ($<20$~nm). Gold chains are only able to transport energy over hundreds of
nanometers with extremely small (2~nm) inter-particle spacing. This is
due to lowering in frequency of the L-mode band as inter-particle
spacing shrinks (gold material suffers from less absorption at
lower frequencies). This argument can be used to explain the increased
energy transfer between gold particles in the so-called
hetero-plasmonic chain waveguide. Effectively a hetero-plasmonic chain can
be treated as superposition of one gold chain and one silver chain: inclusion of silver
particles down-shifts the frequency of L mode of the gold chain, hence
reducing its propagation loss; but the silver chain
waveguide can be adversely affected by presence of gold
particles. A chain waveguide sitting on a dielectric
substrate was also examined, where we noticed degeneracy breakup of
T-mode pair and dominant presence of L mode. At MIR
wavelength, as a demonstration, we numerically showed
that a chain waveguide with plasmonic particles being doped silicon can achieve
deep-subwavelength light guidance, which can be useful for sensing
or thermal applications. 

In general, with realistic plasmonic materials, \emph{subwavelength}
chain modes have very limited propagation lengths. Even for a chain
made of silver, the best plasmonic material
among those we investigated, its subwavelength modes have propagation lengths comparable to the input
light wavelength in free space. Despite heavy losses, chain waveguides
offers more degrees of freedom in engineering modal properties. Although
not attempted in this work, there should be room for
improvements through geometrical optimization. Its deep-subwavelength
mode confinement, locally enhanced field intensity, as well as its
unusual polarization could be useful for various nanophotonic
applications including sensing, nonlinear optics, and efficient coupling of
radiation from quantum emitters, etc.

\section*{Acknowledgement}
The Swedish Research Council (Ventenskapsr{\aa}det, or VR), through project
no. 2016-03911 as well as its Linnaeus center in
Advanced Optics and Photonics (ADOPT), is deeply acknowledged.
The FEM computations were performed on resources provided by the
Swedish National Infrastructure for Computing (SNIC) at PDC - Center
for High Performance Computing at KTH. S. Macinkevicius
is thanked for discussion on semiconductors.


\section*{References}

\newpage

\title{Supplementary data for ``Complex-{$k$} modes of plasmonic chain waveguides''}

\author{M. Yan}

\address{Department of Applied Physics,
 School of Engineering Sciences,
 KTH - Royal Institute of Technology \\
 Isafjordsgatan 22, Kista 16440, Sweden}
\ead{miya@kth.se}
\vspace{10pt}

\begin{figure}[h]
\centering
\includegraphics[width=9cm]{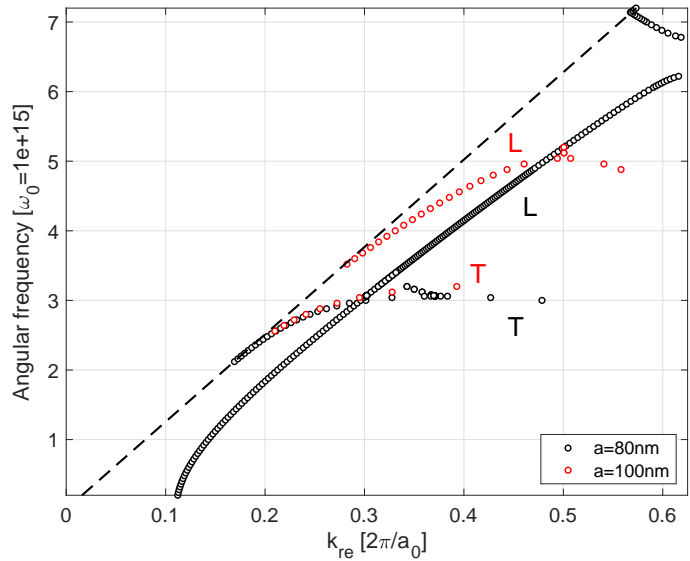}
\caption{Effect of inter-particle gap size (or period) on dispersion
  curves. Dashed line is light line of background medium.}
\label{fig:effect_gap}
\end{figure}

The dispersion curves in Fig.~\ref{fig:effect_gap} show the effect of
gap between nanoparticles. With an increase in gap size (or period
while particle size is not changed), T
mode is not affected much but L mode experiences in general a marked increase in
frequency. Nanoparticles have non-dispersive permittivity with
$\epsilon_m=-6.25$, and diameter $d=60$~nm. Two gap sizes are studied:
20~nm (i.e. $a=80$~nm, presented in main text) and 40~nm
(i.e. $a=100$~nm). Note that the normalization length $a_0$ for wave
number is $a_0=100$~nm in both cases.

\begin{figure}[h]
\centering
\includegraphics[width=9cm]{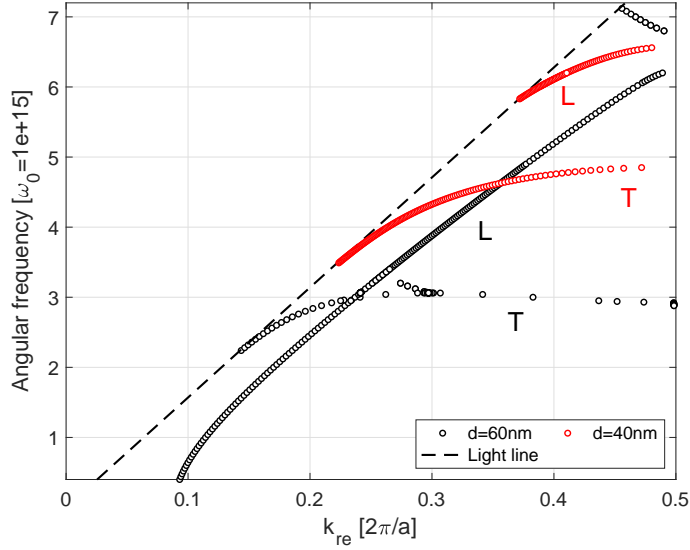}
\caption{Effect of particle size on dispersion curves.}
\label{fig:effect_d}
\end{figure}

The effect of particle size is shown in
Fig.~\ref{fig:effect_d}. Nanoparticles have non-dispersive
permittivity with $\epsilon_m=-6.25$ and fixed period $a=80$~nm. Two
particles sizes are simulated: 60~nm (presented in main text) and
40~nm. Normalization length $a_0$ for wave number is $a_0=80$~nm in
both cases. Smaller particle size increases mode frequencies.

\begin{figure}[h]
\centering
\includegraphics[width=9cm]{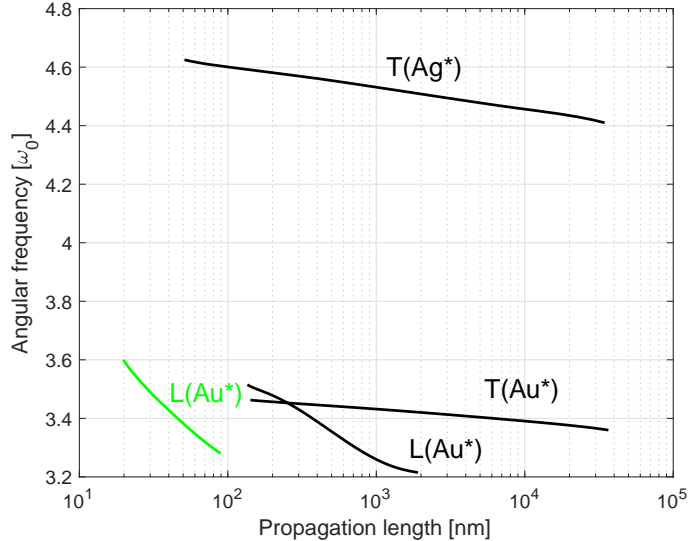}
\caption{Modal propagation losses for the hetero-plasmonic
  waveguide. Black curves are for the case where imaginary parts of gold
  and silver permittivity values are decreased by a factor of
  ten. Green curve is for the case with original metal losses.}
\label{fig:s2im_pLength}
\end{figure}

The propagation lengths of modes supported by the hetero-plasmonic
waveguide are shown in Fig.~\ref{fig:s2im_pLength}. This figure is
complementary to Fig.~7(a) in the main text.

\begin{figure}[h]
\centering
\includegraphics[width=9cm]{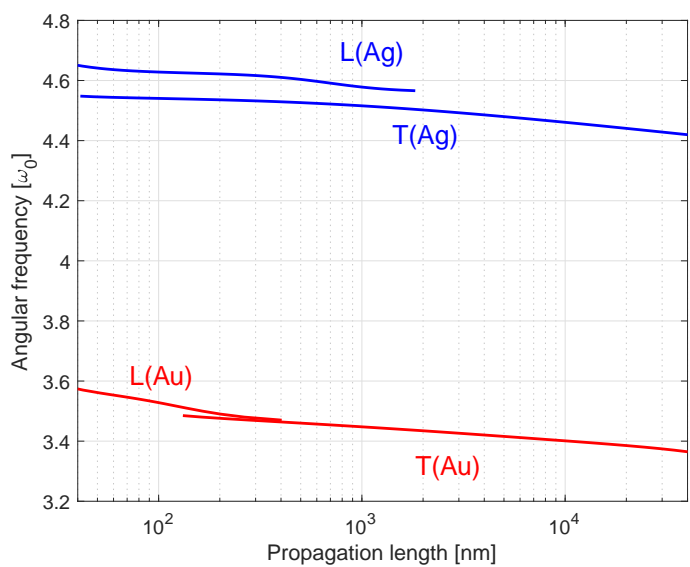}
\caption{Modal propagation losses for two ``homo-plasmonic'' chain
  waveguides. Imaginary parts of gold and silver permittivity values
  are decreased by a factor of ten.}
\label{fig:sim_pLength}
\end{figure}

It is argued in the main text that the hetero-plasmonic chain
waveguide can be treated as two constituent ``homo-plasmonic'' chain
waveguides. The propagation lengths of modes supported by two
corresponding ``homo-plasmonic'' chain waveguides are shown in
Fig.~\ref{fig:sim_pLength}. This figure is complementary to Fig.~7(b)
in the main text.

\begin{figure}[h]
\centering
\includegraphics[width=15cm]{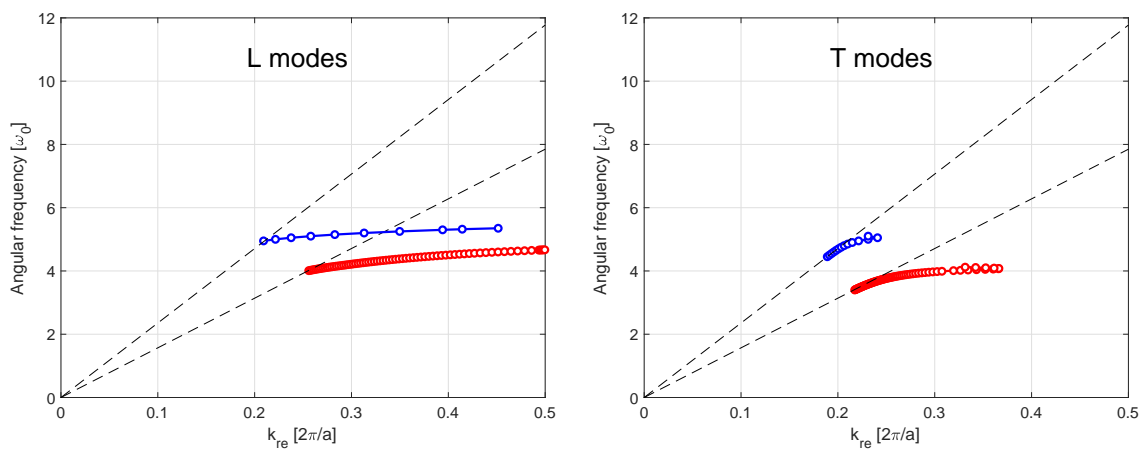}
\caption{Modal dispersion curves for silver-in-air chain waveguide,
  compared to those of silver-in-dielectric chain waveguide. L modes are
  in the left panel and T modes are in the right panel. The blue
  curves are for the air background case, while the red ones are for
  the dielectric background case.}
\label{fig:sia-disp}
\end{figure}

Dispersion curves of T and L modes for a silver-in-air chain waveguide
are shown in Fig.~\ref{fig:sia-disp}. The curves are directly compared
to the case with a dielectric background ($\epsilon_b$=2.25). The
chain waveguide has geometry of $a=80$~nm and $d=60$~nm. Use of air
background shifts the modes to higher frequencies, where silver
experiences higher material loss.

\begin{figure}[h]
\centering
\includegraphics[width=15cm]{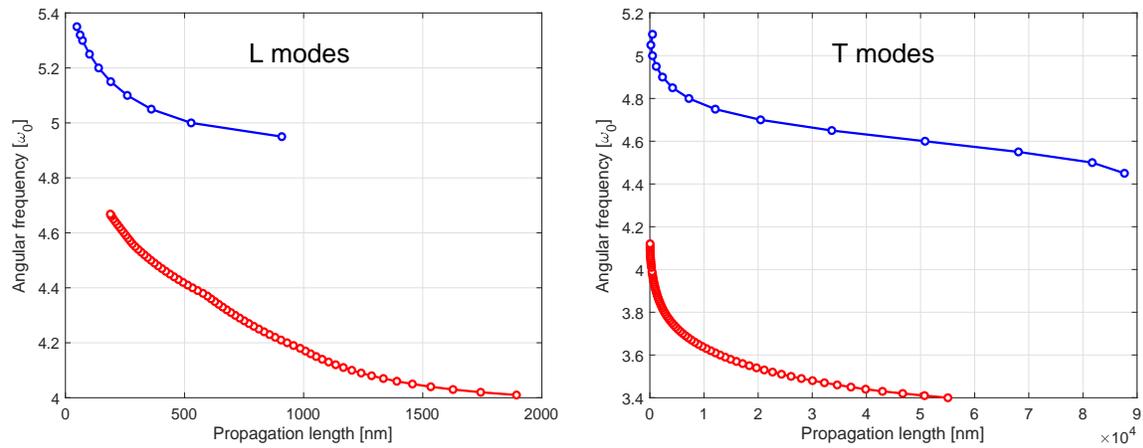}
\caption{Modal propagation lengths for silver-in-air chain waveguide,
  compared to those of silver-in-dielectric chain waveguide. L modes are
  in the left panel and T modes are in the right panel. The blue
  curves are for the air background case, while the red ones are for
  the dielectric background case.}
\label{fig:sia-pLength}
\end{figure}

The corresponding propagation lengths for modes in the silver-in-air
chain waveguide are presented in Fig.~\ref{fig:sia-pLength}. They are
compared directly in the figure against propagation lengths for
modes in a silver-in-dielectric waveguide sharing the same geometry.

\end{document}